\documentclass[mathleft
]{an}
\usepackage{graphicx}
\usepackage[latin1]{inputenc}
\usepackage{times}
\begin{document}

\Pagespan{789}{}
\Yearpublication{}%
\Yearsubmission{}%
\Month{}%
\Volume{}%
\Issue{}%

\title{The Properties of Fossil Groups of Galaxies}

\author{Paul Eigenthaler\thanks {\email{eigenthaler@astro.univie.ac.at} }
\and Werner W. Zeilinger}
\titlerunning{Fossil Groups of Galaxies}
\authorrunning{Paul Eigenthaler, Werner W. Zeilinger}
\institute{
Institut für Astronomie, Universität Wien, Türkenschanzstraße 17, 
A-1180 Wien, Austria}

\received{}
\accepted{}
\publonline{}


\abstract{Numerical  simulations as well  as optical and X-ray  observations over the last  few years  have  shown  that  poor groups of  galaxies can
evolve to what is called a  fossil group. Dynamical friction  as  the driving  process leads to  the coalescence of  individual galaxies in   ordinary
poor groups  leaving  behind nothing  more  than a  central,  massive elliptical galaxy    supposed to  contain  the   merger history   of the   whole
group.  Due to  merging timescales  for less-massive   galaxies and gas cooling timescales of the  X-ray intragroup medium exceeding a Hubble time,  a
surrounding faint-galaxy population having survived  this  galactic cannibalism as well as an  extended X-ray halo similar to  that  found in ordinary
groups, is  expected.  Recent studies suggest   that fossil groups  are  very abundant  and could  be  the progenitors of   brightest cluster galaxies
(BCGs) in  the  centers of rich galaxy clusters. However, only a few objects are known to the literature.  This article  aims to summarize the results
of observational  fossil  group   research over   the  last  few  years  and presents  ongoing   work  by    the authors.  Complementary  to  previous
research, the  SDSS  and  RASS surveys   have been  cross-correlated to  identify new    fossil structures  yielding 34  newly detected   fossil group
candidates. Observations with ISIS  at the 4.2m  William Herschel Telescope  on La Palma  have been carried  out to study  the  stellar populations of
the central ellipticals  of 6 fossil groups.  In  addition multi-object spectroscopy with  VLTs  VIMOS has been performed  to  study the shape of  the
OLF of one fossil system.} 

\maketitle

\section{Introduction}
Redshift surveys have shown  that galaxies  in the  nearby universe are not   randomly distributed in space  but rather  found in  aggregates, so called galaxy  groups  (Geller  \&
Huchra 1983,  Tully  1987).  Basically galaxy groups  can  be  divided  into  three  major classes that differ  mainly in their optical  and physical characteristics: poor  groups,
compact groups, and fossil groups.

Poor or loose groups with a space density of $\sim$$10^{-5}$ Mpc$^{-3}$ (Nolthenius  \& White, 1987) resemble simple multiplets of galaxies, represent
the most common  class  and are simply referred  to  as  galaxy  groups. They are comprised of  approximately  50  members, including   numerous dwarf
galaxies  and take up   typical diameters of  $\sim$1.5  Mpc. Compact groups   with  a  space  density   of  $\sim$1$-$2 $\times$ $10^{-6}$ Mpc$^{-3}$
(Hickson 1982)  are less   abundant  and   typically  made   up    of only   4 to 5  bright  galaxies strongly interacting with    each   other    due
to    their mutual  gravitational  attraction    causing    tidal  forces    to form   tails  around   closely interacting    members. Since    today,
about  1000 compact     groups  have    been   classified   (Hickson    1997).   Fossil groups       are  characterised      through a      dominating
central bright elliptical        galaxy   comparable   to   cD    galaxies  in      clusters   surrounded   by      a    diffuse      and     extended
X-ray   halo together  with      a  significant   fainter galaxy    population. Recent studies have   estimated space densities of  fossil systems  in
the local universe to  be $\sim$1$-$4 $\times$  $10^{-6}$ Mpc$^{-3}$, comparable  to  HCGs. Besides   the fact that   the vast majority   of  galaxies
in the universe    appear to be   located  in   groups,  these systems   play   also a  key  role  in our understanding  of   galaxy evolution.  Since
the  masses   of   groups     are  comparatively      small   ($\sim$$10^{13}$    M$_{\odot}$)   with       respect  to     the     richer    clusters
($\sim$$10^{14}$  M$_{\odot}$),   the   virial  theorem   implies  lower    values  for    the  group  velocity      dispersions    favoring    galaxy
interactions leading  to the coalescence of individual  galaxies, galaxy merging. 

Numerical  simulations have  shown that   this cannibalism can   proceed  as long   as  a single,   massive elliptical galaxy   remains  as
final  product  of    multiple  merging events   (Barnes 1989)  on  a  timescale of   a few  gigayears.  The  physical   origin  behind the merging of
galaxies is explained   by dynamical   friction  of the  group  member galaxies with  the group dark-matter  halo and  acts   not  for all    types of
galaxies  similarly effective. According to the description  of dynamical  friction by  Chandrasekhar (1943), the friction  force is   strongly   mass
dependent leading  to   longer   merging    timescales  (see Binney   \&  Tremaine p.429)   for  less  massive  galaxies  which therefore should  have
survived this  galactic   cannibalism   up to  now. Moreover,     the    hot   X-ray    gas,      as   found  in    $\sim$50\%     of   all      known
galaxy     groups  (not  correlated  with  individual    group   members     but    rather  diffusely     distributed    following     the      global
gravitational    group    potential  (Mulchaey  2000))   is also expected  to be observed    in  this   state    of   hierarchical  galaxy   evolution
since  gas temperatures    and  electron densities of a   typical  intragroup   medium    with $T$ $\sim$ 10$^{7}$ K  and    $n_{e}$ $\sim$  10$^{-3}$
cm$^{-3}$  suggest    gas   cooling     timescales    ${\tau   _{{\rm{cool}}}}   \propto   {n_e}^{-1}{T_g}^{1/2}$  exceeding    the    age  of     the
universe. Thus,     one  claims   that \textit   {fossil   groups}   represented by     a central    bright, massive  elliptical galaxy surrounded  by
a  diffuse   and extended   X-ray halo  and  a   fainter galaxy  population are the remnants     of  what was  initally a  poor   group   of  galaxies
that has been  transformed  to this  old, \textit {fossil}  stage of    galaxy evolution  in  low    density environments with   compact groups acting
as likely  way  station in  this evolution.  If this scenario is correct   then the relative space densities of these three different types of systems
give information on their transformation rates. Moreover, Jones et    al. (2003)  assume  that fossil  systems constitute of  probably    8-20\%    of
all   systems with     comparable X-ray   luminosity  ($\ge$10$^{42}$       ergs   s$^{-1}$),   thus acting    also very   likely as   the  site    of
formation  of brightest cluster galaxies (BCGs) before the infall   into clusters. Similar optical luminosities of the central ellipticals   in fossil
groups compared  to that of  BCGs support this  idea.  Therefore    fossils  play  a   crucial   role  in   our   understanding  of  galaxy  evolution
in   low-density environments  and  the   physical connection between  groups  and clusters respectively. 

The definition  of a  fossil system as proposed by Jones et al.  (2003)  was the  first real attempt to  attribute exact observational charateristics  to this class.   According to
their influental paper a fossil system is a spatially  extended X-ray source with an X-ray luminosity from diffuse, hot  gas of $L_{X,\textrm{bol}} \ge 10^{42}$ h$_{50}^{-2}$  ergs
s$^{-1}$. The optical  counterpart is a bound system  of galaxies  with $\Delta m_{1,2} \ge 2.0$  mag,  where  $\Delta m_{1,2}$ is the absolute total magnitude  difference in   the
Johnson $R$ band between the two brightest galaxies  in the system   within half  the projected  virial radius.  No upper  limit is  given  for the X-ray  luminosity
or  temperature. The   ideas behind  this  classification scheme are  simple.   The  lower  limit in   X-ray luminosity excludes normal   bright elliptical  galaxies  exhibiting  a
hot coronal gas  component.  The magnitude   gap  of $\Delta  m_{1,2}   \ge 2.0$  mag for the  two brightest galaxies  of  the  group ensures   that  one  single elliptical  galaxy
dominates  the system.   Studies of luminosity functions   of galaxy groups  have   shown  that   this  threshold   acts  as   a  good   criterion  to   distinguish  between   poor
groups  and  fossil  aggregates  since  observations have shown that such  a high magnitude gap is  extremely unlikely to occur in  ordinary groups or clusters (Beers et al. 1995).

The first fossil group identified, RX J1340.6+4018, was  published in Nature (Ponman   et al. 1994). Since  then, few  other objects  have been assigned   to this
class resulting   in  a   catalogue of   15  fossil  systems  summarized by  Mendes de  Oliveira et  al.(2006).

\begin            {table} [h!]
\begin            {center}
\caption          {\label {15}Fossil galaxy groups as summarized by Mendes de Oliveira et al. (2006). RX denotations refer to ROSAT X-ray sources.}
\begin {footnotesize}
\begin            {tabular}{lcclcl}
\hline \hline
\multicolumn{1}{c}{Name} &   $\alpha_{2000}$  &  $\delta_{2000}$  &  \multicolumn{1}{c}{$z$}   & L$_{X,bol}$$^{a}$  \\
\hline
NGC 1132                 &     02 52 51.8     &     -01 16 29     &            0.023           &     1.9            \\
RX~J0454.8-1806          &     04 54 52.2     &     -18 06 56     &            0.031           &     1.9            \\
ESO 306- G 017           &     05 40 06.7     &     -40 50 11     &            0.036           &     129            \\
RX J1119.7+2126          &     11 19 43.7     &     +21 26 50     &            0.061           &     1.7            \\
RX~J1159.8+5531          &     11 59 51.4     &     +55 32 01     &            0.081           &     22             \\
CL 1205+44               &     12 05 53.7     &     +44 29 46     &            0.59            &     180            \\
RX~J1256.0+2556          &     12 56 03.4     &     +25 56 48     &            0.232           &     61.            \\
RX~J1331.5+1108          &     13 31 30.2     &     +11 08 04     &            0.081           &     5.9            \\
RX~J1340.6+4018          &     13 40 33.4     &     +40 17 48     &            0.171           &     25             \\
RX~J1416.4+2315          &     14 16 26.9     &     +23 15 32     &            0.137           &     220.           \\
RX~J1552.2+2013          &     15 52 12.5     &     +20 13 32     &            0.136           &     63             \\
NGC 6034                 &     16 03 32.1     &     +17 11 55     &            0.034           &     0.75           \\
NGC 6482                 &     17 51 48.8     &     +23 04 19     &            0.013           &     2.17           \\
RX~J2114.3-6800          &     21 14 20.4     &     -68 00 56     &            0.130           &     20             \\ 
RX~J2247.4+0337          &     22 47 29.1     &     +03 37 13     &            0.199           &     41             \\ 
\hline
\end              {tabular} \\
\end {footnotesize}
\begin            {flushleft}
$^{a}$ bolometric X-ray luminosity taken in 10$^{42}$ h$_{50}^{-2}$ ergs s$^{-1}$ \\
\end              {flushleft}
\end              {center}
\end              {table}

\begin            {figure} [h!]
\begin            {center}
\includegraphics  [width=238pt] {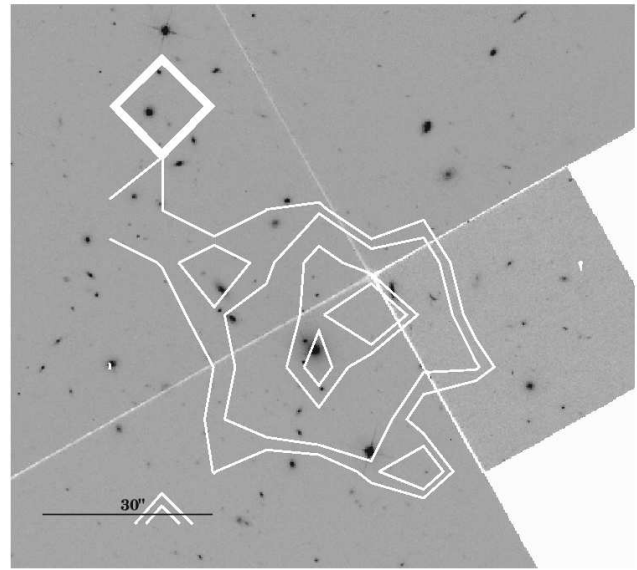}
\caption          {\label{ulmergroup}F702W \textit{HST} image of Cl1205+44, the most  distant fossil group known. Chandra contours in the  range 0.5-8keV indicate the diffuse and extended X-ray  gas
component. North is up, east is to the left. Taken from Ulmer et al. (2005).}
\end              {center}
\end              {figure}

However,  not   all listed   objects represent    what is  typically referred   to as     fossil  group   since more  massive systems   with  velocity
dispersions in the  range   of 600    km s$^{-1}$, thus  more appropriate to   galaxy clusters, are   also present (Cypriano  et  al. 2006, Mendes  de
Oliveira et al.  2006). More recently,  Santos et al. (2007) have   cross-correlated optical and X-ray data from the  Sloan Digital Sky Survey (SDSS)
(Adelman-McCarthy  et  al. 2008) and the ROSAT All Sky  Survey (RASS) (Voges et al.  1999) respectively to identify new fossil group candidates. Their
findings comprise a list of 34  candidates,  ranging  up to $z$ $=$ 0.489. The   majority  of  all known 49  systems ($\sim$86\%) resides   in the    northern
celestial  hemisphere   whereas   only  7   objects      ($\sim$14\%) have been   identified  in  the  southern one.  The  nearest  fossil group   was
identified with the elliptical NGC 6482 at a  redshift of $z$ $=$ 0.013  (Khosroshahi et  al. 2004)  while the most  distant fossil aggregate was  found  at
$z$ $=$ 0.59 combining HST, Chandra  and  XMM-Newton data (Ulmer et al. 2005, see Fig. \ref{ulmergroup}).

\begin            {figure} [h!]
\begin            {center}
\includegraphics  [width=238pt] {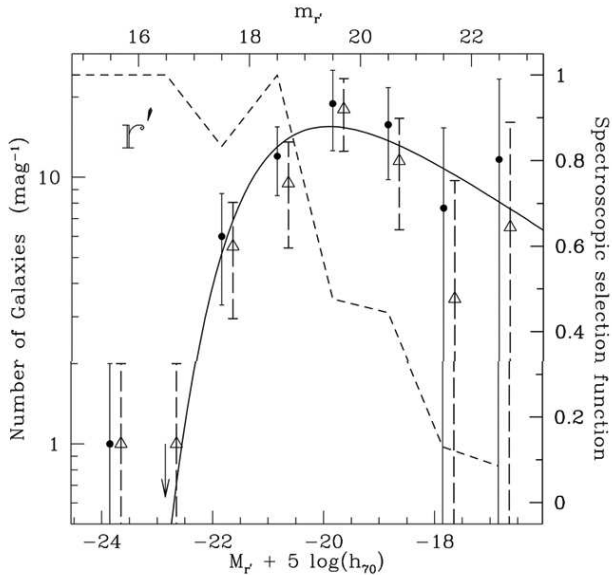}
\caption          {\label {lumfunc}Luminosity function of the fossil  cluster RX J1552.2+2013 in the SDSS  $r^{\prime}$ band. Solid circles show the completeness-corrected  number of spectroscopically
confirmed members  while triangles  show photometrically  determined members  estimated through  number counts  and statistical  background subtraction. The dotted line represents a
selection function of the spectroscopic sample while the continuous line shows the best Schechter fit to the spectroscopic sample excluding the brightest galaxy. A clear dip in the
LF can be found around $M_{r'}$ $=$ $-18$. Taken from Mendes de Oliveira et al. (2006).}
\end              {center}
\end              {figure}

\section{The properties of fossil groups}

Since the catalogue of known fossil aggregates is only made up of a few objects, an analysis of the optical or X-ray properties of fossil groups has only been carried out for a few
systems  leaving the formation process of this special class of objects still unclear. This section highlights some of the results that have been obtained over the last few years.

\subsection {The  luminosity function of  fossil  galaxy groups} 
The  optical luminosity   function  (OLF)  of  fossil  groups exhibits  a lack   of bright  galaxies (by    definition since   fossil  systems     are
characterised  as low-density  environments with   no bright   galaxies  besides    the central elliptical).   Thus,  the   bright  end  of   the  OLF
of   these  systems  is   unusual  compared  to    what     is  normally    observed    in  poor     groups,  with    too  few    $L$$^{*}$   galaxies
present.  Few  effort   has   been  made so     far  to  study   the  shape  of   the   OLF  at  fainter  magnitudes.      Existing  studies  focusing
on the  fainter part  of the  OLF of  fossil galaxy  groups  have  only investigated  three  fossil      systems. Surprisingly,  all  these aggregates
are  rather    classified as  fossil clusters  than  groups    due    to their   relatively    large  velocity dispersions   (623  km    s$^{    -1}$,
584   km  s$^{-1}$, and 565 km s$^{-1}$)(Cypriano  et  al.    2006, Mendes de    Oliveira   et        al. 2006,  Mendes  de   Oliveira  et         al.
2009). Interestingly, besides    the expected    magnitude  gap  around   $L$$^{*}$ galaxies,           a   dip in the     OLF   of   RX  J1552.2+2013
at     the transition to  dwarf galaxies at $M_{r'}=-18$   has been   found  (see Fig.  \ref{lumfunc}). Such  dips in   the     OLFs   have   been observed      in
several   systems, mostly      dense,    dynamically     well-evolved  aggregates   such    as     X-ray  emitting   cD      clusters    (Valotto   et
al.      2004)    or compact     groups      of      galaxies    (Hunsberger      et   al.    1996),   suggesting  that   there     may   be      more
than       one mechanism  driving     the     depletion     of      galaxies       in      these  different    environments. Assuming     the  merger
scenario  hypothesis   from  hierarchical  galaxy    evolution   models,    the    lack    of  $L$$^{*}$     galaxies   is  explained    by  dynamical
friction,  resulting   in   a deceleration   of  the   orbital  velocity    of   a   galaxy     which   loses     its   kinetic energy        to   the
surrounding   dark    matter particles,     and      subsequent    merging.     The    same     physical      processes     cannot        be efficient
enough   to       reproduce    the  comparatively   small         number     of   low-mass  galaxies    around $M_{R}=-18$.      One  explanation  for
these missing galaxies could  be    a succession  of tidal     encounters (Gnedin     2003). If   this scenario  is   correct,     one  would      not
expect     such a      dip in    the luminosity  functions  for   less  rich    galaxy   aggregates  than  the     observed  fossil  cluster,    since
the  galaxy density and    thus  the tidal  stripping      efficiency   is   much      lower    in  these       systems.   However,  RX   J1416.4+2315
and  RXJ1340.6+4018 do  not show   any evidence  of    a dip  in  the  OLF around    $M_{R}=-18$,  suggesting  that  the  scenario  of   fossil  group
formation is far more complicated  than expected.  More studies  focusing   on multi-object spectroscopy of   the faint-galaxy population in    fossil
groups or fossil   systems in general is therefore needed.

\subsection {Masses and mass-to-light ratios of fossil groups}  
The symmetric and  regular X-ray  emission  together  with the  lack  of $L$$^{*}$  galaxies and  the   large early-type fraction  observed in  fossil
groups suggest that these   systems  are old,   evolved systems.    Thus,    from a   dynamical   point of   view,   fossil  groups    should  present
relaxed, virialized galaxy aggregates.   The  dynamical   properties   of the faint-galaxy  population  of  fossil  systems studied so  far show  that
the radial velocity distributions exhibit far too  high velocity dispersions to  be considered  as remnants of what  was initially a  poor   group  of
galaxies and are  fairly  well fit   by  a gaussian  velocity   distribution,  indicating  relaxed  systems. It is feasible to estimate  the
masses  of fossil  groups from  the gas density and temperature  profiles  of the extended X-ray   gas component assuming  the gas  distribution to be
spherically symmetric  and in hydrostatic equilibrium. Then the total gravitational mass is given by

\begin{equation}
{M_{{\rm{grav}}}}( < r) =  - \frac{{kT(r)r}}{{G\mu {m_{\rm{p}}}}}\left[ {\frac{{d\ln \rho (r)}}{{d\ln r}} + \frac{{d\ln T(r)}}{{d\ln r}}} \right]
\end{equation}

where $\mu$ is the mean molecular weight  and $m_{p}$ the proton mass. X-ray  studies confirm the high  masses derived  from the
dynamical  state of  these systems, showing  that the intragroup  medium of a  number  of fossil  groups is similar  to those of  galaxy clusters with
temperatures exceeding even 4 keV  for some aggregates. These results  are inconsistent with the idea  that fossil groups are the  remains of galactic
cannibalism in ordinary groups.  Complementary,  Yoshioka   et  al.  (2004)   has   shown  that   the   M/L ratios of    four   fossil  groups    reach   up
to  $M/L_{B}=1100$ M$_{\odot}$/L$_{\odot}$ being at least  one order    of  magnitude  higher than the  typical    M/L ratios for   groups  and      clusters
of  comparable   mass supporting the idea that it  is  unlikely that   these systems    are   the  remnants  of regular groups  and   clusters.  Other
recent studies    derive  much lower  values   of  M/L,  however (e.g.  Sun  et  al.   2004),  showing that a   general  formation scenario for fossil
systems is far from  understood and further observations  in  both optical and  X-rays are  needed to get reliable  mass estimates for a
higher number of fossil aggregates and to see if all fossil structures resemble cluster-like systems rather than groups.

\subsection {Surface brightness profiles of the central ellipticals}  
The  shape   of the    surface  brightness    profiles of    ellipticals  are known   to  depend   on  the   formation histories    of these
systems and can  thus    act  as indicators  for    different  imaginable    evolution tracks.  More precisely, dissipationless    merger  simulations
from   Naab \&  Burkert (2003)  have   shown  that  unequal-mass mergers  lead   to fast   rotating,   discy   ellipticals, while  equal mass  mergers
produce slowly  rotating, pressure supported  systems.  Major mergers between bulge dominated systems result in boxy ellipticals, independent of   the
mass ratio  while merger remnants  that  subsequently accrete gas  are  always discy.  More recently, it   has been shown  that the  mergers of spiral
galaxies alone cannot reproduce the   kinematic  and  photometric properties  of very   massive  ellipticals  (Naab, Khochfar  \& Burkert   2006)  nor
can   they reproduce the observed correlation   between isophotal   shapes and  the luminosity    of ellipticals.   Khoshroshahi et  al. (2006)   have
investigated    the profile shapes of   the  central  ellipticals in   seven fossil  systems   with  the conclusion   that the  isophote    shapes  of
ellipticals    in fossil groups   are  different  compared  to  the brightest  central galaxies  in non-fossil  systems,   especially  rich  clusters.
Luminous  elliptical    galaxies in    non-fossil  systems  do  not    present  discy  isophotes    (Kormendy   \&  Bender 1996). In    contrast, the
brightest ellipticals    in   fossil groups show  discy and  boxy  isophotes  in similar   proportions. If  the  central  ellipticals in  groups  have
indeed formed   via the merging  of all  other major galaxies,  then   some of these merging   events  would have  been  gas rich,  since  groups  are
known  to  consist of  a large  spiral     fraction.  The discy     charateristics   of   central   ellipticals     in fossil   groups   would    then
be    consistent    with    numerical   simulations, that   discy isophotes   result from    gas-rich  mergers.    The difference    in the   observed
isophote    shapes of    fossil  group     central galaxies    and  the  brightest  cluster     galaxies does    not   rule    out  the       possible
evolutionary   link  of   infalling   pre-processed    merged ellipticals from groups    in clusters, however.  This  could  still  be the case     if
these ellipticals  have  undergone later   gas-free  mergers within   the   cluster environment.    Moreover,  about 40\% of    the  brightest cluster
galaxies  exhibit  at  least  one  secondary   nucleus, strongly  supporting the   idea of  late mergers   within the   cluster environment.   Further
investigation of the   isophote shape characteristics of central ellipticals  in   fossil  groups will  help to   confirm  or question  the   existing
results  and  will  shed  more    light on    the  evolution of    fossil group central ellipticals and possible  connection to the brightest  cluster
galaxies.
 
\subsection {Scaling relations in fossil groups}  
The first study of the   scaling properties of   fossil groups  compared  to  ordinary groups  and  clusters was  carried  out by Khoshroshahi  et al.
(2007). Based  upon Chandra  X-ray observations  as well  as optical  imaging and  spectroscopy this  study comprised the information of seven  fossil
groups showing  regular,  symmetric   X-ray emission,   indicating  no   recent  mergers.   Scaling relations   focusing  on  the total  gravitational
mass,   X-ray temperature,   X-ray luminosity,    group velocity   dispersion and    the  optical   luminosity were studied. For a given  optical
luminosity of  the  whole group, fossil  systems turn out   to be more  X-ray luminous than  non-fossil  groups. Focusing  on the  $L_{X}$$-$$T_{X}$
relation,  fossils   cannot    be  distinguished   from   ordinary  groups   or    clusters,  however.   Fossils   also  show   that for a given group
velocity  dispersion,  X-ray  luminosity  and  temperature  are  higher  compared  to  non-fossil  systems.  The  $M_{X}$$-$$T_{X}$ relation  suggests
that fossils are hotter,  for a given  total gravitational mass, consistent with an early formation epoch.

\section          {The search for new fossil group candidates}
Complementary to the  previous work on  fossil  groups as  highlighted in section 2, the  attempt  to identify new  fossil structures was   made by the authors. This
undertaking seems reasonable  since the space  densities of fossil  groups determined so  far suggest that   these systems are  as abundant as Hickson
compact groups while the actual catalogue  comprises only a negligible fraction compared to HCGs. The $\Lambda$CDM concordance model ($\Omega_{m}$ $=$
0.3, $\Omega_{\Lambda}$ $=$ 0.7, $H_{0}$ $=$ 70 km s$^{-1}$) was used throughout this work. Luminosity distances $D_{L}$ and angular diameter distances
$D_{A}$ have been   calculated via the   \emph{Cosmology Calculator} (Wright   2006). In order   to identify new   fossil group candidates,  the Sloan
Digital  Sky   Survey  and      Rosat  All Sky Survey   have been  cross-correlated  to  shortlist
candidates  of interest  that  meet  the classification  criteria  as proposed by Jones  et al. (2003). The  query  was defined via   Structured Query
Language  (SQL). A similar approach has   been  carried out by Santos  et al.  (2007) focusing on  different selction  criteria, however.   The  query
presented here  focuses  on  intrinsically bright ${{M_{{r'}}} \le   - 21}+5\log  h$,  red  ${g' - r' >  0.8}$ galaxies  relating to  ellipticals. The
lower limit in absolute magnitude was ensured by calculating the corresponding apparent  magnitudes via luminosity distances $D_{L}$ derived from  the
SDSS spectroscopic redshifts of the central ellipticals

\begin{equation}
m <  M + 25 + 5\log \left[ {\frac{{D_L }}{{h^{ - 1} {\rm{Mpc}}}}} \right] + A + K(z)
\end{equation}

accounting for galactic extinction $A$ and $K$-correction too. Only galaxies with an entry  in the SDSS ROSAT  table were considered for the   further
shortlisting of  fossil group   candidates. The  associated X-ray  component  had  to be  extended with  a ROSAT  extent parameter  of at  least   one
arcsecond.  This value  does  not  reflect the  true  extent   of the  X-ray  source  but gives  the  excess  over the   ROSAT point-spread  function.
Furthermore the distance of the ROSAT source had to be less than 100kpc from the central elliptical accounting also for the ROSAT position error.  The
SQL code is given in appendix A. The resulting sample was finally shortlisted  via the ${2^{{\rm{mag}}}}$ criterion forcing all systems
to have  no galaxies with $\Delta  m_{1,2} \ge 2.0$ mag in  the  SDSS $r^{\prime}$ band within  one-half virial  radius. In  contrast to the work   of
Santos et al.  (2007) who use a  constant value of $0.5  h^{-1}$ Mpc for half the virial radius, the description of Evrard et al. (1996) 

\begin{equation}
r_{{\rm{vir}}}  = 1.945 \cdot \left( {\frac{T}{{10{\rm{ kev}}}}} \right)^{1/2} \left( {1 + z} \right)^{ - 3/2}  \cdot h^{ - 1} {\rm{ Mpc}} \\ 
\end{equation}

as  used  by Jones  et  al. (2003)   was  applied here   assuming  a lower  limit   of the  X-ray  temperature of  fossil   groups  of  0.7  keV.  The
${2^{{\rm{mag}}}}$  criterion  was  accounted  for  via an  SDSS  SQL function   (see  appendix   A).    The  remaining   objects  were  checked  for
morphology. Five galaxies  were identified  as spirals  while  three objects have  been  found in  the vicinity  of bright stars.  These systems  have
been  excluded from the sample  leaving  a final list   of  34  previously nondetected  fossil group    candidates. The    query also  reidentified  
objects listed      in the Santos et  al. (2007) sample.  However, many objects  could not  be  reidentified since the   procedure  used  here  excludes  all
galaxies within  half the  virial    radius  not taking     into   account   possible   background  galaxies,   thus    eliminating possible    fossil
group  candidates that would pass     the ${2^{{\rm{mag}}}}$  criterion    if   group  memberships  were  known. This indicates   that   there   might
be  far   more    fossil  candidates  in   our query    when  accounting     also  for     group memberships      via  photometric   redshifts.  Table
\ref{fossilcoordinates}   gives  coordinates  as   well  as    SDSS  and      ROSAT  identifications of   the  new   fossil   group candidates.  Table
\ref{fossilproperties} presents  the most important properties  of  the fossil  group   central ellipticals as   well as  the  associated  ROSAT X-ray
component. Fig.  5  shows  a colour-magnitude  diagram of  the  newly found fossil group central   ellipticals as  well   as  the
Santos  et  al.   (2007)  sample. The   symbol size indicates  the     prominence  of   the   associated    X-ray  source,   visualizing   the   ROSAT
extent/$\Delta$   ratio   as   presented   in  Table  \ref{fossilproperties}.  Arrows  point  to  galaxies that   have been  observed with the William
Herschel Telescope (WHT) and the Very   Large Telescope (VLT).  In  order to    get   a glimpse  on  the   optical   luminosity function     (OLF)  of
the    newly    detected fossil   group candidates,   SDSS photometric   and   spectroscopic redshifts, when  available,  have  been used  to estimate
group memberships  and construct  an  OLF  for all  systems. Some   of  the OLFs confirm the   lack of faint galaxies    as  previously  observed   by
Mendes de Oliveira et   al. (2006) while most   of the    systems don't   show a   dip in the OLF at fainter magnitudes. Fig.   \ref{gap}    presents the  photometrically   selected
OLF  of  RX J1520.9+4840  from  the    new    list  of    fossil aggregates. A  clear dip   in  the  OLF  is   visible    around    $M_{i'}=-18$.   It
has   to  kept    in  mind, however,   that    this     result is   based   upon photometric    redshifts which  exhibit  far   larger   errors   than
spectroscopic   measurements necessitating the use   of multi-object   spectroscopy of the  faint   galaxy population  to confirm the lack of faint galaxies in  fossil galaxy groups.

\section          {Observations}
\begin            {figure} [h!]
\begin            {center}
\includegraphics  [width=238pt] {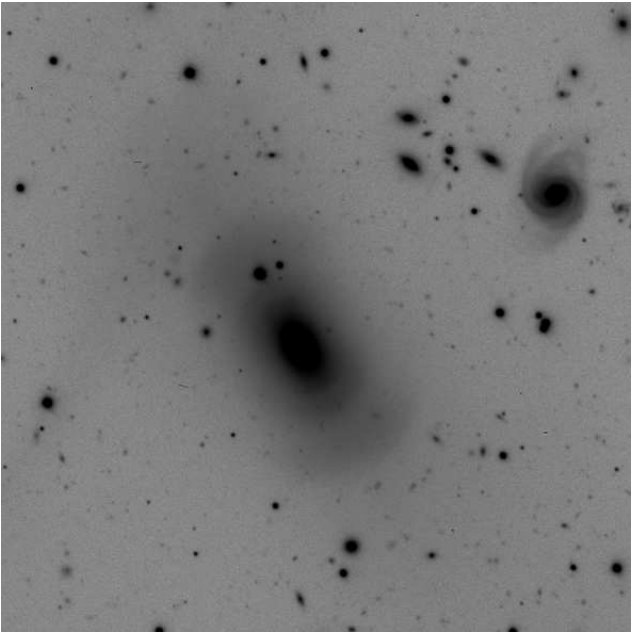}
\caption          {\label {shells}VIMOS PRE-IMG of the fossil group candidate RX J1548.9+0851 from the Santos et al. (2007) sample. The bright spiral shows a redshift difference of $\Delta z \sim
0.008$  to the central elliptical and doesn't violate  the 2$^{\rm{mag}}$ criterion. Shells can be  clearly seen  around the elliptical which suggest multiple past  merging
events  that would confirm the status of RX J1548.9+0851 as a fossil group. North is up, east is to the left. The field of view is $\sim$2 arcmin on a side.}
\end              {center}
\end              {figure}

\begin            {figure} [h!]
\begin            {center}
\includegraphics  [width=238pt] {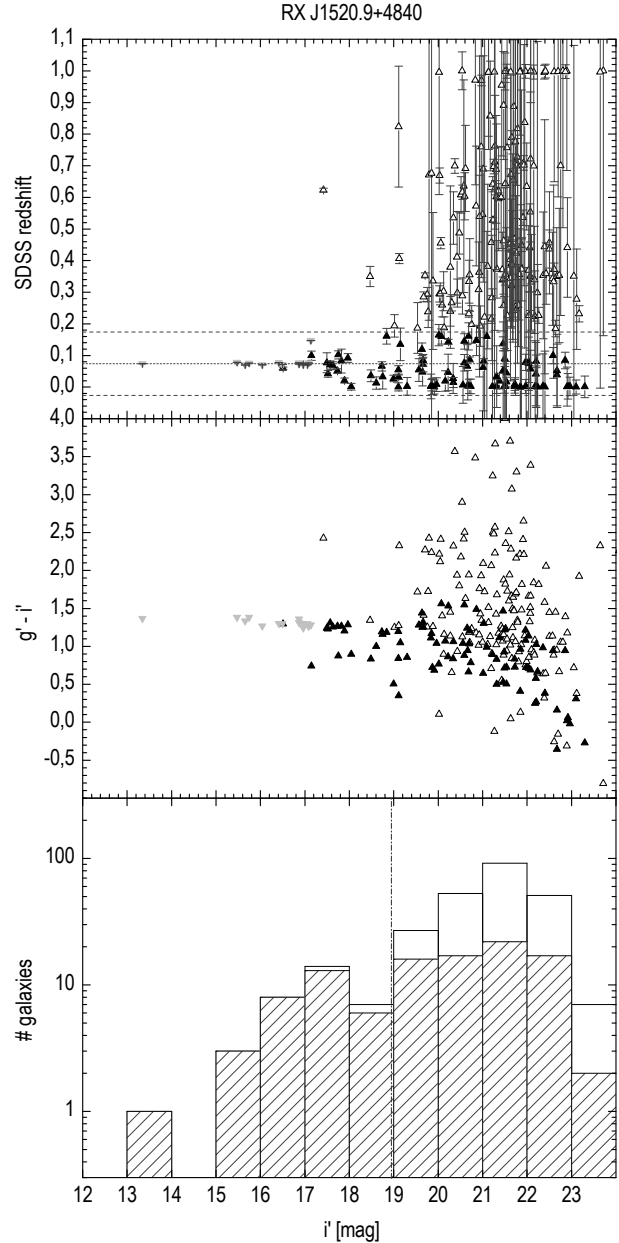}
\caption          {\label {gap}Photometrically determined optical luminosity function of   the fossil group candidate RX J1520.9+4840  from the new sample. All objects  in the SDSS  classified
as galaxies   within one-half  virial radius   are plotted.  Upper  panel:  Open and   black triangles  represent  SDSS  photometric redshifts   while gray triangles  indicate SDSS
spectroscopic redshifts. The dotted  line  indicates the   spectroscopic redshift  $z$    of the central  elliptical,  while    dashed lines at   $z\pm \Delta  z$ ($\Delta  z=0.1$;
SDSS photometric  redshift uncertainty) indicate borders    for group   membership.   Mid  panel: Colour-Magnitude   Diagram  of the   investigated galaxy   population.    Galaxies
exhibiting adequate photometric  redshifts for group  membership follow  a  red sequence as  observed  in ordinary  groups. Lower   panel:  Photometrically-determined luminosity
function. Blank    histograms show  all galaxies   while dashed  ones indicate  objects within  $z\pm \Delta z$.  The dash  dotted line  corresponds to  $M_{i'} =  -18$. A  dip  in
the photometrically selected OLF around $M_{i'} = -18$ is  clearly visible.}
\end              {center}
\end              {figure}

Table \ref{observations} summarizes  the observations at    the WHT and      VLT that have    been carried    out  to complement the  existing data on
fossil groups. Long-slit spectra have  been  obtained with  the  ISIS  spectrograph  at   the    WHT to    measure  Lick-IDS  indices (Worthey et  al.
1994)   of    the  central  ellipticals from both    the   new sample  as   well  as fossil  groups from   the  literature. These  measurements   will
yield  radial profiles of    ages and metallicities     of   the central     elliptical stellar  populations    that  are  supposed   to contain   the
merger history   of     the  whole group.  It   is   then   feasible  to   study    how uniform   the     process   of     coalescence   occured    in
these systems  and    if  fossil     group    central  ellipticals   are indeed   the progenitors   of    brightest  cluster    galaxies   (BCGs)   or
show  similar  stellar   populations  as  found  in non BCGs.  In addition,   VIMOS observations    in service  mode   have    been carried  out    at
VLT  UT3   Melipal   to  study   the   OLF of     one fossil    aggregate. Spectroscopic  targets   have   been selected
via SDSS      magnitudes  to    determine  the    OLF  down    to $i^{\prime}\sim  20$  corresponding   to $M_{i'}\sim   -17$.  This magnitude   limit
ensures to  identify  a  dip  in the OLF   around $M_{i'}\sim     -18$ if  present. Fig. \ref{shells} shows   the   VIMOS  pre-image of  the    central
elliptical  of  the  fossil     group candidate  RX J1548.9+0851.  The  image reveals  shells  around  the  central  elliptical  that  would   confirm
the idea of the system being  the remainder of multiple merging  events. The aquired data   from the WHT  and the  VLT  will  shed more  light on
the  formation process of  fossil  groups of  galaxies  complementing the few optical  studies that  have been carried out so far.

\begin            {table*} [h!]
\begin            {center}
\caption          {\label{observations} Fossil systems observed with the WHT and the VLT.}
\begin            {tabular}{lcccccc}
\hline \hline
\multicolumn{1}{c}{galaxy} &  $\alpha_{2000}$   &   $\delta_{2000}$   &    exposure [s]    &  \multicolumn{2}{c}{resolution [\AA \space pix$^{-1}$]}    &    date       \\
\hline                                                                                                                                                                    
                           &                    &                     &                    &                             &                              &               \\
\multicolumn{7}{c}{\raisebox{0.5ex}[-0.5ex]{\textit{William Herschel Telescope}$^{a}$}}                                                                                 \\
\hline                                                                                                                                                                    
                           &                    &                     &                    &            blue             &             red              &               \\
\hline                                                                                                                                                                    
RX J0752.7+4557            &    07  52  44.2    &    +45  56  57.4    &   3$\times$ 3000   &            0.86             &             0.26             &  20.12.2008   \\
RX J0844.9+4258            &    08  44  56.6    &    +42  58  35.7    &   3$\times$ 3000   &            0.86             &             0.26             &  20.12.2008   \\
NGC 1132                   &    02  52  51.8    &    -01  16  28.8    &   4$\times$ 2700   &            0.86             &             0.26             &  20.12.2008   \\
RX J1548.9+0851            &    15  48  55.9    &    +08  50  44.4    &   3$\times$ 2600   &            0.86             &             0.93             &  28.04.2009   \\
RX J1520.9+4840            &    15  20  52.3    &    +48  39  38.6    &   3$\times$ 2600   &            0.86             &             0.93             &  28.04.2009   \\
RX J1152.6+0328            &    11  52  37.6    &    +03  28  21.8    &   3$\times$ 2700   &            0.86             &             0.93             &  28.04.2009   \\
\hline                                                                                                                                                                    
                           &                    &                     &                    &                             &                              &               \\
\multicolumn{7}{c}{\raisebox{0.5ex}[-0.5ex]{\textit{Very Large Telescope}$^{b}$}}                                                                                       \\
\hline                                                                                                                                                                    
RX J1548.9+0851            &    15  48  55.9    &    +08  50  44.4    &        4500        &                  \multicolumn{2}{c}{0.51}                  &  26.05.2009   \\
\hline                                                                                           
\end              {tabular} \\
\vspace {0.2cm}
\begin  {flushleft}
$^{a}$  Observations carried out in Visitor Mode.                                            \\
$^{b}$  Observations carried out in Service Mode. The date refers to the completion of the last Observing Block (OB) for the given target. \\
\end    {flushleft}
\end              {center}
\end              {table*}

\begin            {table*} [h!]
\centering
\caption          {\label {fossilcoordinates}Fossil group candidates: ROSAT and SDSS identifications.}
\begin            {tabular}{p{3cm}p{3cm}p{2.5cm}p{2.5cm}p{3cm}}
\hline\hline 
    \multicolumn{1}{c}{ID Number}  &  \multicolumn{1}{c}{ROSAT Name}     &  \multicolumn{1}{c}{SDSS $\alpha_{2000}$$^{a}$} &  \multicolumn{1}{c} {SDSS $\delta_{2000}$$^{a}$} &   \multicolumn{1}{c}   {SDSS objid}      \\  
    \multicolumn{1}{c}{(1)}        &  \multicolumn{1}{c}{(2)}            &  \multicolumn{1}{c}{(3)}                        &  \multicolumn{1}{c}         {(4)}                &   \multicolumn{1}{c}       {(5)}         \\  
\hline                                                                                                                                                                                                                       
      01\dotfill                   &  \centering{RXJ0159.8-0850} &  \centering{01  59  49.3}               &    \centering{-08  49  58.8}             &  \multicolumn{1}{c}{587727884161581256}  \\  
      02\dotfill                   &  \centering{RXJ0725.7+3741} &  \centering{07  25  41.3}               &    \centering{+37  40  27.7}             &  \multicolumn{1}{c}{587737826206482926}  \\  
      03\dotfill                   &  \centering{RXJ0730.3+3728} &  \centering{07  30  20.4}               &    \centering{+37  27  40.8}             &  \multicolumn{1}{c}{587725775602450780}  \\  
      04\dotfill                   &  \centering{RXJ0801.0+3603} &  \centering{08  00  56.8}               &    \centering{+36  03  23.6}             &  \multicolumn{1}{c}{587728905564258619}  \\  
      05\dotfill                   &  \centering{RXJ0825.9+0415} &  \centering{08  25  57.8}               &    \centering{+04  14  48.3}             &  \multicolumn{1}{c}{587732702851170810}  \\  
      06\dotfill                   &  \centering{RXJ0826.9+3108} &  \centering{08  26  57.6}               &    \centering{+31  08  04.9}             &  \multicolumn{1}{c}{587732469849325635}  \\  
      07\dotfill                   &  \centering{RXJ0909.9+3106} &  \centering{09  09  53.3}               &    \centering{+31  06  03.2}             &  \multicolumn{1}{c}{588016878295318532}  \\  
      08\dotfill                   &  \centering{RXJ1005.8+1058} &  \centering{10  05  50.7}               &    \centering{+10  58  11.7}             &  \multicolumn{1}{c}{587734949665243313}  \\  
      09\dotfill                   &  \centering{RXJ1006.1+0710} &  \centering{10  06  08.3}               &    \centering{+07  10  29.5}             &  \multicolumn{1}{c}{587732579379904650}  \\  
      10\dotfill                   &  \centering{RXJ1030.5+1416} &  \centering{10  30  34.8}               &    \centering{+14  15  41.8}             &  \multicolumn{1}{c}{587735349637349534}  \\  
      11\dotfill                   &  \centering{RXJ1041.8+5303} &  \centering{10  41  47.6}               &    \centering{+53  03  41.4}             &  \multicolumn{1}{c}{587733081343656013}  \\  
      12\dotfill                   &  \centering{RXJ1055.1+4246} &  \centering{10  55  06.6}               &    \centering{+42  45  24.4}             &  \multicolumn{1}{c}{588017626148634721}  \\  
      13\dotfill                   &  \centering{RXJ1056.1+0252} &  \centering{10  56  06.6}               &    \centering{+02  52  13.5}             &  \multicolumn{1}{c}{587726033315299465}  \\  
      14\dotfill                   &  \centering{RXJ1107.4+1245} &  \centering{11  07  24.2}               &    \centering{+12  44  20.0}             &  \multicolumn{1}{c}{588017567629509093}  \\  
      15\dotfill                   &  \centering{RXJ1115.9+0130} &  \centering{11  15  51.9}               &    \centering{+01  29  55.1}             &  \multicolumn{1}{c}{587728307491897590}  \\  
      16\dotfill                   &  \centering{RXJ1128.6+3530} &  \centering{11  28  34.4}               &    \centering{+35  30  14.0}             &  \multicolumn{1}{c}{587739305286238362}  \\  
      17\dotfill                   &  \centering{RXJ1152.6+0328} &  \centering{11  52  37.6}               &    \centering{+03  28  21.8}             &  \multicolumn{1}{c}{587726033858330762}  \\  
      18\dotfill                   &  \centering{RXJ1211.1+3520} &  \centering{12  11  08.3}               &    \centering{+35  19  58.8}             &  \multicolumn{1}{c}{587739304216232052}  \\  
      19\dotfill                   &  \centering{RXJ1327.1+0212} &  \centering{13  27  01.0}               &    \centering{+02  12  19.5}             &  \multicolumn{1}{c}{587726015078465728}  \\  
      20\dotfill                   &  \centering{RXJ1349.9+4217} &  \centering{13  49  51.1}               &    \centering{+42  16  47.8}             &  \multicolumn{1}{c}{588017604154687567}  \\  
      21\dotfill                   &  \centering{RXJ1352.0+6105} &  \centering{13  51  58.3}               &    \centering{+61  04  19.0}             &  \multicolumn{1}{c}{588011219135889619}  \\  
      22\dotfill                   &  \centering{RXJ1353.4+4424} &  \centering{13  53  20.3}               &    \centering{+44  24  19.8}             &  \multicolumn{1}{c}{588298662507184284}  \\  
      23\dotfill                   &  \centering{RXJ1402.8+3431} &  \centering{14  02  46.6}               &    \centering{+34  31  08.0}             &  \multicolumn{1}{c}{587739131880603717}  \\  
      24\dotfill                   &  \centering{RXJ1431.3-0054} &  \centering{14  31  21.2}               &    \centering{-00  53  44.3}             &  \multicolumn{1}{c}{588848898855469518}  \\  
      25\dotfill                   &  \centering{RXJ1450.2+4134} &  \centering{14  50  08.3}               &    \centering{+41  33  59.8}             &  \multicolumn{1}{c}{588017116130246662}  \\  
      26\dotfill                   &  \centering{RXJ1453.6+0359} &  \centering{14  53  38.5}               &    \centering{+03  59  33.4}             &  \multicolumn{1}{c}{587726101487747355}  \\  
      27\dotfill                   &  \centering{RXJ1501.3+5455} &  \centering{15  01  18.0}               &    \centering{+54  55  18.3}             &  \multicolumn{1}{c}{588011101565354082}  \\  
      28\dotfill                   &  \centering{RXJ1520.9+4840} &  \centering{15  20  52.3}               &    \centering{+48  39  38.6}             &  \multicolumn{1}{c}{587735666921898155}  \\  
      29\dotfill                   &  \centering{RXJ1539.8+4143} &  \centering{15  39  51.4}               &    \centering{+41  43  25.4}             &  \multicolumn{1}{c}{587733397568356378}  \\  
      30\dotfill                   &  \centering{RXJ1540.4+3622} &  \centering{15  40  23.0}               &    \centering{+36  21  56.6}             &  \multicolumn{1}{c}{587736751928770630}  \\  
      31\dotfill                   &  \centering{RXJ1624.7+3727} &  \centering{16  24  43.4}               &    \centering{+37  26  42.4}             &  \multicolumn{1}{c}{587735666392105305}  \\  
      32\dotfill                   &  \centering{RXJ1653.1+3909} &  \centering{16  53  07.8}               &    \centering{+39  08  53.1}             &  \multicolumn{1}{c}{588007005270966692}  \\  
      33\dotfill                   &  \centering{RXJ1717.1+2931} &  \centering{17  17  06.9}               &    \centering{+29  31  21.1}             &  \multicolumn{1}{c}{587729408084935169}  \\  
      34\dotfill                   &  \centering{RXJ2139.5-0722} &  \centering{21  39  28.5}               &    \centering{-07  21  46.6}             &  \multicolumn{1}{c}{587726878878728621}  \\  
\hline                                                                                                                                                                                                                       
\end              {tabular}                                                                                                                                                                                              \\  
\vspace {0.2cm}                                                                                                                                                                                                              
$^{a}$ Right ascension is given in hours, minutes, and seconds and declination is given in degrees, arcminutes, and arcseconds.                                                                                          \\  
\end              {table*}

\begin            {table*} [h!]
\centering
\caption          {\label {fossilproperties}Fossil group candidates: Properties of the central elliptical and the associated X-ray component.}
\begin            {tabular}{p{2.2cm}p{1cm}p{1cm}p{1cm}p{1cm}p{1cm}p{1cm}p{0.9cm}p{0.9cm}p{0.9cm}}
\hline\hline 
                               &                                                         \multicolumn{4}{c}{central elliptical galaxy}                                                                             &                                                         &                    \multicolumn{4}{c}{\textit{ROSAT} X-ray properties}                                                                                             \\  
                               &                                     &  \multicolumn{1}{c}{i$^{\prime}$$^{a}$} & \multicolumn{1}{c}{g$^{\prime} - i^{\prime}$$^{a}$}  & \multicolumn{1}{c}{M$_{i^{\prime}}$$^{a}$} & \multicolumn{1}{c}{\textonehalf $r_{{\rm{vir}}}$$^{b}$} & \multicolumn{1}{c}{$\Delta$$^{c}$} & \multicolumn{1}{c}{extent$^{c}$}  &                                           & \multicolumn{1}{c}{$L_{X}$ (0.5-2keV)$^{d}$}  \\  
 \multicolumn{1}{c}{ID Number} &  \multicolumn{1}{c}{redshift$^{a}$} &  \multicolumn{1}{c}{[mag]}              & \multicolumn{1}{c}{[mag]}                            & \multicolumn{1}{c}{[mag]}                  & \multicolumn{1}{c}{[arcmin]}                            & \multicolumn{1}{c}{[arcsec]}       & \multicolumn{1}{c}{[arcsec]}      & \multicolumn{1}{c}{extent/$\Delta$$^{c}$} & \multicolumn{1}{c}{[ergs s$^{-1}$]}           \\  
 \multicolumn{1}{c}{(1)}       &  \multicolumn{1}{c}{(2)}            &  \multicolumn{1}{c}{(3)}                & \multicolumn{1}{c}{(4)}                              & \multicolumn{1}{c}{(5)}                    & \multicolumn{1}{c}{(6)}                                 & \multicolumn{1}{c}{(7)}            & \multicolumn{1}{c}{(8)}           & \multicolumn{1}{c}{(9)}                   & \multicolumn{1}{c}{(9)}                       \\  
\hline                                                                                                                                                                                                                                                                                                                                                                                                                                                
01  \dotfill                   &  \centering{0.405}                  &  \centering{17.29}                      & \centering{2.00}                                     & \centering{-24.37}                         & \centering{0.680}                                       & \multicolumn{1}{r}{12.8}           & \multicolumn{1}{r}{44}            & \multicolumn{1}{r}{ 3.44}                 & \multicolumn{1}{c}{8.42E+44}                  \\  
02  \dotfill                   &  \centering{0.425}                  &  \centering{17.34}                      & \centering{2.58}                                     & \centering{-24.56}                         & \centering{0.647}                                       & \multicolumn{1}{r}{42.8}           & \multicolumn{1}{r}{56}            & \multicolumn{1}{r}{ 1.31}                 & \multicolumn{1}{c}{1.84E+44}                  \\  
03  \dotfill                   &  \centering{0.200}                  &  \centering{17.58}                      & \centering{1.69}                                     & \centering{-22.04}                         & \centering{1.411}                                       & \multicolumn{1}{r}{21.8}           & \multicolumn{1}{r}{40}            & \multicolumn{1}{r}{ 1.83}                 & \multicolumn{1}{c}{6.29E+43}                  \\  
04  \dotfill                   &  \centering{0.287}                  &  \centering{16.13}                      & \centering{2.23}                                     & \centering{-24.30}                         & \centering{0.970}                                       & \multicolumn{1}{r}{35.5}           & \multicolumn{1}{r}{47}            & \multicolumn{1}{r}{ 1.33}                 & \multicolumn{1}{c}{3.85E+44}                  \\  
05  \dotfill                   &  \centering{0.225}                  &  \centering{15.81}                      & \centering{2.01}                                     & \centering{-24.31}                         & \centering{1.249}                                       & \multicolumn{1}{r}{21.4}           & \multicolumn{1}{r}{21}            & \multicolumn{1}{r}{ 0.98}                 & \multicolumn{1}{c}{1.51E+44}                  \\  
06  \dotfill                   &  \centering{0.209}                  &  \centering{15.26}                      & \centering{1.92}                                     & \centering{-24.43}                         & \centering{1.346}                                       & \multicolumn{1}{r}{52.1}           & \multicolumn{1}{r}{76}            & \multicolumn{1}{r}{ 1.46}                 & \multicolumn{1}{c}{1.25E+44}                  \\  
07  \dotfill                   &  \centering{0.272}                  &  \centering{16.57}                      & \centering{1.20}                                     & \centering{-23.65}                         & \centering{1.026}                                       & \multicolumn{1}{r}{ 9.2}           & \multicolumn{1}{r}{22}            & \multicolumn{1}{r}{ 2.39}                 & \multicolumn{1}{c}{6.65E+44}                  \\  
08  \dotfill                   &  \centering{0.162}                  &  \centering{15.15}                      & \centering{1.68}                                     & \centering{-24.10}                         & \centering{1.755}                                       & \multicolumn{1}{r}{ 7.4}           & \multicolumn{1}{r}{85}            & \multicolumn{1}{r}{11.45}                 & \multicolumn{1}{c}{4.58E+43}                  \\  
09  \dotfill                   &  \centering{0.202}                  &  \centering{17.25}                      & \centering{1.35}                                     & \centering{-22.39}                         & \centering{1.394}                                       & \multicolumn{1}{r}{48.9}           & \multicolumn{1}{r}{22}            & \multicolumn{1}{r}{ 0.45}                 & \multicolumn{1}{c}{1.89E+43}                  \\  
10  \dotfill                   &  \centering{0.317}                  &  \centering{16.23}                      & \centering{2.31}                                     & \centering{-24.49}                         & \centering{0.875}                                       & \multicolumn{1}{r}{26.1}           & \multicolumn{1}{r}{61}            & \multicolumn{1}{r}{ 2.34}                 & \multicolumn{1}{c}{2.07E+44}                  \\  
11  \dotfill                   &  \centering{0.187}                  &  \centering{16.52}                      & \centering{1.71}                                     & \centering{-22.59}                         & \centering{1.517}                                       & \multicolumn{1}{r}{28.2}           & \multicolumn{1}{r}{13}            & \multicolumn{1}{r}{ 0.46}                 & \multicolumn{1}{c}{1.97E+43}                  \\  
12  \dotfill                   &  \centering{0.371}                  &  \centering{18.57}                      & \centering{1.48}                                     & \centering{-23.05}                         & \centering{0.744}                                       & \multicolumn{1}{r}{ 7.2}           & \multicolumn{1}{r}{17}            & \multicolumn{1}{r}{ 2.36}                 & \multicolumn{1}{c}{1.00E+44}                  \\  
13  \dotfill                   &  \centering{0.236}                  &  \centering{17.60}                      & \centering{1.28}                                     & \centering{-22.27}                         & \centering{1.189}                                       & \multicolumn{1}{r}{ 6.1}           & \multicolumn{1}{r}{19}            & \multicolumn{1}{r}{ 3.12}                 & \multicolumn{1}{c}{1.03E+45}                  \\  
14  \dotfill                   &  \centering{0.420}                  &  \centering{17.93}                      & \centering{2.46}                                     & \centering{-23.12}                         & \centering{0.655}                                       & \multicolumn{1}{r}{15.8}           & \multicolumn{1}{r}{ 9}            & \multicolumn{1}{r}{ 0.57}                 & \multicolumn{1}{c}{1.19E+44}                  \\  
15  \dotfill                   &  \centering{0.352}                  &  \centering{17.19}                      & \centering{2.03}                                     & \centering{-24.26}                         & \centering{0.786}                                       & \multicolumn{1}{r}{22.5}           & \multicolumn{1}{r}{57}            & \multicolumn{1}{r}{ 2.54}                 & \multicolumn{1}{c}{7.30E+44}                  \\  
16  \dotfill                   &  \centering{0.402}                  &  \centering{17.59}                      & \centering{2.52}                                     & \centering{-23.59}                         & \centering{0.685}                                       & \multicolumn{1}{r}{15.6}           & \multicolumn{1}{r}{12}            & \multicolumn{1}{r}{ 0.77}                 & \multicolumn{1}{c}{3.02E+44}                  \\  
17  \dotfill                   &  \centering{0.081}                  &  \centering{14.25}                      & \centering{1.38}                                     & \centering{-23.06}                         & \centering{3.600}                                       & \multicolumn{1}{r}{27.4}           & \multicolumn{1}{r}{71}            & \multicolumn{1}{r}{ 2.59}                 & \multicolumn{1}{c}{1.08E+43}                  \\  
18  \dotfill                   &  \centering{0.136}                  &  \centering{15.14}                      & \centering{1.58}                                     & \centering{-23.52}                         & \centering{2.097}                                       & \multicolumn{1}{r}{35.8}           & \multicolumn{1}{r}{52}            & \multicolumn{1}{r}{ 1.45}                 & \multicolumn{1}{c}{4.11E+43}                  \\  
19  \dotfill                   &  \centering{0.260}                  &  \centering{16.26}                      & \centering{2.16}                                     & \centering{-24.11}                         & \centering{1.075}                                       & \multicolumn{1}{r}{40.5}           & \multicolumn{1}{r}{58}            & \multicolumn{1}{r}{ 1.43}                 & \multicolumn{1}{c}{2.30E+44}                  \\  
20  \dotfill                   &  \centering{0.289}                  &  \centering{17.67}                      & \centering{1.92}                                     & \centering{-22.77}                         & \centering{0.964}                                       & \multicolumn{1}{r}{ 8.2}           & \multicolumn{1}{r}{19}            & \multicolumn{1}{r}{ 2.30}                 & \multicolumn{1}{c}{1.08E+44}                  \\  
21  \dotfill                   &  \centering{0.323}                  &  \centering{17.32}                      & \centering{2.34}                                     & \centering{-23.47}                         & \centering{0.858}                                       & \multicolumn{1}{r}{23.2}           & \multicolumn{1}{r}{ 6}            & \multicolumn{1}{r}{ 0.26}                 & \multicolumn{1}{c}{3.61E+43}                  \\  
22  \dotfill                   &  \centering{0.152}                  &  \centering{17.03}                      & \centering{1.37}                                     & \centering{-21.48}                         & \centering{1.873}                                       & \multicolumn{1}{r}{24.1}           & \multicolumn{1}{r}{ 9}            & \multicolumn{1}{r}{ 0.37}                 & \multicolumn{1}{c}{1.23E+43}                  \\  
23  \dotfill                   &  \centering{0.175}                  &  \centering{17.14}                      & \centering{1.31}                                     & \centering{-22.61}                         & \centering{1.619}                                       & \multicolumn{1}{r}{ 6.7}           & \multicolumn{1}{r}{10}            & \multicolumn{1}{r}{ 1.49}                 & \multicolumn{1}{c}{1.28E+43}                  \\  
24  \dotfill                   &  \centering{0.403}                  &  \centering{17.49}                      & \centering{2.70}                                     & \centering{-23.80}                         & \centering{0.684}                                       & \multicolumn{1}{r}{43.1}           & \multicolumn{1}{r}{52}            & \multicolumn{1}{r}{ 1.21}                 & \multicolumn{1}{c}{2.27E+44}                  \\  
25  \dotfill                   &  \centering{0.157}                  &  \centering{14.89}                      & \centering{1.65}                                     & \centering{-24.02}                         & \centering{1.818}                                       & \multicolumn{1}{r}{35.9}           & \multicolumn{1}{r}{54}            & \multicolumn{1}{r}{ 1.50}                 & \multicolumn{1}{c}{2.24E+43}                  \\  
26  \dotfill                   &  \centering{0.370}                  &  \centering{16.82}                      & \centering{2.35}                                     & \centering{-24.24}                         & \centering{0.746}                                       & \multicolumn{1}{r}{36.1}           & \multicolumn{1}{r}{42}            & \multicolumn{1}{r}{ 1.16}                 & \multicolumn{1}{c}{2.33E+44}                  \\  
27  \dotfill                   &  \centering{0.339}                  &  \centering{17.24}                      & \centering{1.83}                                     & \centering{-23.63}                         & \centering{0.818}                                       & \multicolumn{1}{r}{ 7.3}           & \multicolumn{1}{r}{15}            & \multicolumn{1}{r}{ 2.05}                 & \multicolumn{1}{c}{1.54E+44}                  \\  
28  \dotfill                   &  \centering{0.074}                  &  \centering{13.35}                      & \centering{1.37}                                     & \centering{-24.12}                         & \centering{3.933}                                       & \multicolumn{1}{r}{31.4}           & \multicolumn{1}{r}{71}            & \multicolumn{1}{r}{ 2.26}                 & \multicolumn{1}{c}{4.35E+43}                  \\  
29  \dotfill                   &  \centering{0.119}                  &  \centering{14.83}                      & \centering{1.55}                                     & \centering{-23.96}                         & \centering{2.404}                                       & \multicolumn{1}{r}{10.3}           & \multicolumn{1}{r}{12}            & \multicolumn{1}{r}{ 1.17}                 & \multicolumn{1}{c}{7.06E+42}                  \\  
30  \dotfill                   &  \centering{0.231}                  &  \centering{16.53}                      & \centering{1.76}                                     & \centering{-23.35}                         & \centering{1.213}                                       & \multicolumn{1}{r}{27.7}           & \multicolumn{1}{r}{11}            & \multicolumn{1}{r}{ 0.40}                 & \multicolumn{1}{c}{2.63E+43}                  \\  
31  \dotfill                   &  \centering{0.199}                  &  \centering{16.89}                      & \centering{1.79}                                     & \centering{-23.99}                         & \centering{1.417}                                       & \multicolumn{1}{r}{12.1}           & \multicolumn{1}{r}{13}            & \multicolumn{1}{r}{ 1.08}                 & \multicolumn{1}{c}{2.97E+43}                  \\  
32  \dotfill                   &  \centering{0.147}                  &  \centering{15.70}                      & \centering{1.64}                                     & \centering{-23.14}                         & \centering{1.945}                                       & \multicolumn{1}{r}{33.0}           & \multicolumn{1}{r}{19}            & \multicolumn{1}{r}{ 0.58}                 & \multicolumn{1}{c}{1.40E+43}                  \\  
33  \dotfill                   &  \centering{0.278}                  &  \centering{16.95}                      & \centering{1.68}                                     & \centering{-22.12}                         & \centering{1.003}                                       & \multicolumn{1}{r}{ 4.2}           & \multicolumn{1}{r}{30}            & \multicolumn{1}{r}{ 7.08}                 & \multicolumn{1}{c}{6.92E+44}                  \\  
34  \dotfill                   &  \centering{0.410}                  &  \centering{17.58}                      & \centering{2.50}                                     & \centering{-23.88}                         & \centering{0.672}                                       & \multicolumn{1}{r}{26.3}           & \multicolumn{1}{r}{20}            & \multicolumn{1}{r}{ 0.76}                 & \multicolumn{1}{c}{1.53E+44}                  \\  
\hline                                                                                                                                                                                                                                                                                                                                                                                                                                                
\end              {tabular}                                                                                                                                                                                                                                                                                                                                                                                                                       \\  
\vspace {0.2cm}
\begin  {flushleft}
$^{a}$ Redshifts and magnitudes are taken from the SDSS DR 6. \\
$^{b}$ Half  the virial  radius given  in arcminutes.  The virial  radius was  estimated via  $r_{{\rm{vir}}}  = 1.945 \cdot \left( {\frac{T}{{10{\rm{
kev}}}}} \right)^{1/2} \left( {1 + z} \right)^{ - 3/2}  \cdot  h^{ - 1}  {\rm{  Mpc}}$ as used in  Jones et al. (2003).  For the X-ray temperature   a
lower limit of  0.7 keV was assumed.  \\
$^{c}$ ROSAT $\Delta$ and extent prameters. $\Delta$ gives the distance of the X-ray source to the central elliptical. Extent gives the excess of  the
detected X-ray source over the ROSAT  PSF.  The ratio extent/$\Delta$ indicates  the  prominence of the   X-ray source with  larger  values indicating
more extended  sources close to  the  central elliptical.\\
$^{d}$      ROSAT   X-ray      luminosities.   ROSAT      countrates   have      been    converted      to   X-ray      fluxes   via      the   tool
PIMMS    (Mukai, 1993)  assuming  a  Raymond-Smith   model  with  2keV   and a  metallicity    of   $Z=0.4  Z_{\sun}$.   HI  column  densities of  the
Leiden/Argentine/Bonn (LAB)  Survey of  Galactic HI (Kalberla et al. 2005) were taken into account to determine extinction-free fluxes.\\
\end  {flushleft}
\end              {table*}

\begin            {figure*} [h!]
\begin            {center}
\includegraphics  [width=420pt] {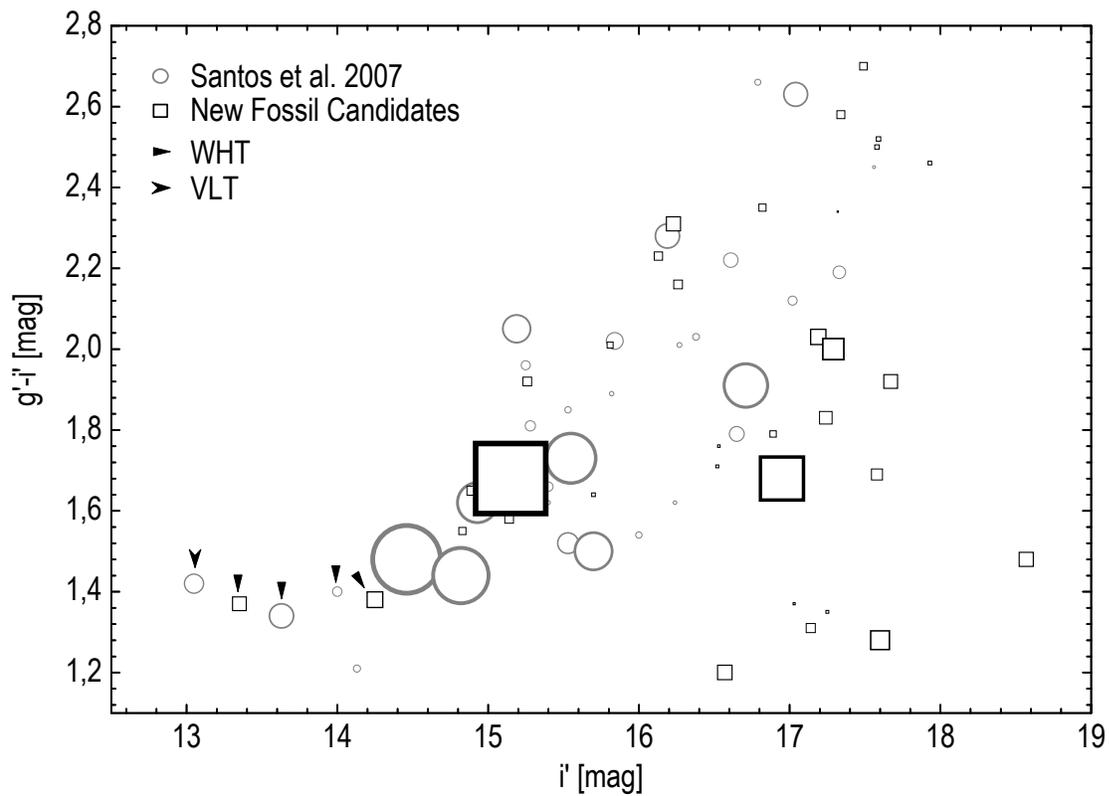}
\caption          {\label {cmd}Colour-magnitude diagram for the central  elliptical galaxies of fossil structures identified  in the SDSS. Circles show the  list of Santos et al. (2007)  while
squares represent new structures  identfied by the authors.  The arrows point to  objects that have been  observed with the WHT  and the VLT by  the authors. The size  of the symbols
indicates the extent/$\Delta$ ratio as presented in Table \ref{fossilproperties}.}
\end              {center}
\end              {figure*}

\acknowledgements
Paul Eigenthaler is supported by the University of Vienna in the frame of the Initiative Kolleg (IK) \textit{The Cosmic Matter Circuit} I033-N.

\appendix

\begin {onecolumn}
\section{SQL Query}

\textit{
SQL query used within SDSS DR6 to select the fossil group candidates listed in Table \ref{fossilcoordinates}.} 

\begin {footnotesize}
\begin{verbatim}
SELECT p.objid, s.z, p.ra, p.dec, p.r, p.g-p.r, r.delta, r.cps, r.cpserr, r.extent
FROM specobj s, photoobj p, rosat r
WHERE s.bestobjid= p.objid  
\end{verbatim}
\begin {flushleft}
\textit{
Selection of objects that have been spectroscopically classified as galaxies with $z\le0.5$ and show an entry  in the SDSS ROSAT  table.}
\end {flushleft}
\begin{verbatim}
AND s.specclass = 2 
AND p.objid= r.objid        
AND s.z BETWEEN 0.0 AND 0.50
\end{verbatim}
\begin {flushleft}
\textit{                            
Selection of bright (${{M_{{r'}}} \le    - 21}+5\log  h$), red (${g^{\prime}-r^{\prime}   > 0.8}$) galaxies via the  relation:}  $m < M +  25 +  5\log
\left( {{D_L}[{\rm{Mpc]}} \cdot h} \right)  + A  + K(z)$\footnote {$D_{L}$ and $K(z)$ were implemented  via third order polynomials that were  fit  to
the $D_{L}$ and $K(z)$ data derived  from the \textit{Cosmology  Calculator} and SDSS  for redshifts up to  $z=0.5$. The  \texttt{kcorr\_r}  entry  in
the   SDSS  photoz   table   was not   directly    used   in the    query  since  these   values   are  based  upon photometric redshifts    which can
strongly    deviate   from   spectroscopic  redshifts  and   would     lead    to    wrong    results.  Therefore  $K(z)$  was determined   based upon
galaxies with   ${g^{\prime} -  r^{\prime} > 0.8}$ that show hardly any difference in photometric and spectroscopic redshifts.}
\end {flushleft}                          

\begin{verbatim}
AND p.r<                    
-21.0            
+ 25 + 5*LOG10 ((-0.15958 + 4290.00033 *s.z + 3255.86186 *s.z*s.z - 1009.42877 *s.z*s.z*s.z) *0.7)
- 0.00554 + 1.31479 *s.z - 1.08771 *s.z*s.z + 5.09347 *s.z*s.z*s.z 
+ p.extinction_r    
AND p.g - p.r > 0.8 
\end{verbatim}
                    
\begin {flushleft}
\textit{
Selecting galaxies with an extended X-ray  source within 100kpc taking into account  the ROSAT position error. \space \footnote {The  polynomial  gives the angular scale as a  function of
redshift based upon angular diameter distances $D_{A}$.}
}
\end {flushleft}
\begin{verbatim}
AND r.extent >= 1 
AND r.delta <= 100/(0.00717 + 20.42378 *s.z - 21.97056 *s.z*s.z + 11.06756 *s.z*s.z*s.z) + r.poserr
\end{verbatim}

\begin {flushleft}
\textit{
Identifying non-fossil systems. All objects within half the  virial radius that are photometrically classified as galaxies  and less than 2 mag fainter than the  central elliptical
are counted. \footnote {This query has  been performed for all galaxies shortlisted  via the previous criteria. \texttt{ra, dec}  give the coordinates of the central  galaxy in
decimal degrees while \texttt{radius} presents half the virial radius in arcminutes. \texttt{m1} stands for the $r'$ band magnitude of the central galaxy.}
}
\end {flushleft}

\begin{verbatim}
SELECT count(*)-1 
FROM  photoObj p, dbo.fGetNearbyObjEq(ra, dec, radius) n
                        
WHERE p.objID = n.objID 
AND p.type = 3      
AND p.r - m1 < 2.0  
\end{verbatim}      
\end{footnotesize}  
\end {onecolumn}
\end{document}